\documentclass[aip,reprint,showpacs,floatfix,amsmath,amssymb]{revtex4-1}

\usepackage{graphicx, color}
\usepackage{amsmath,amssymb,wasysym}
\usepackage{verbatim}

\newcommand{\slfrac}[2]{\left.#1\middle/#2\right.}

\newcommand{\figOne}{
\begin{figure}[t]
	\centering
	\includegraphics[width=3.5in]{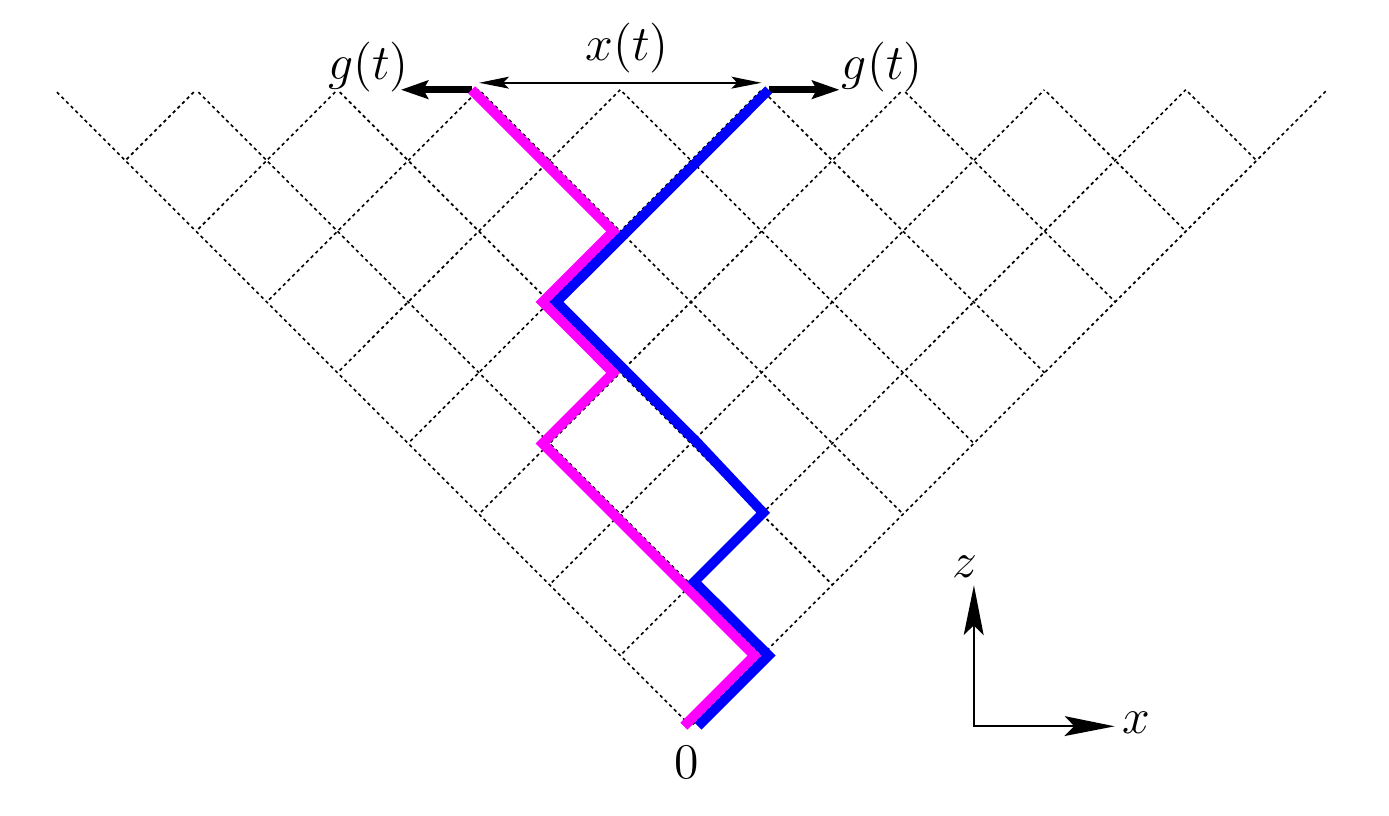}

	\caption{ \label{fig:1} Schematic diagram of the model. The strands of 
	the DNA are shown by thick solid lines. The end monomers of the strands are pulled along 
	$x$ direction with a periodic force $g(t) = G \left| \sin (\omega t) \right|$. The 
	separation between the end monomers, $x(t)$, follows the external force $g(t)$ with a lag.
	}

\end{figure}
}

\newcommand{\figTwo}{
\begin{figure}[t]
	\centering
	\includegraphics[width=3.0in]{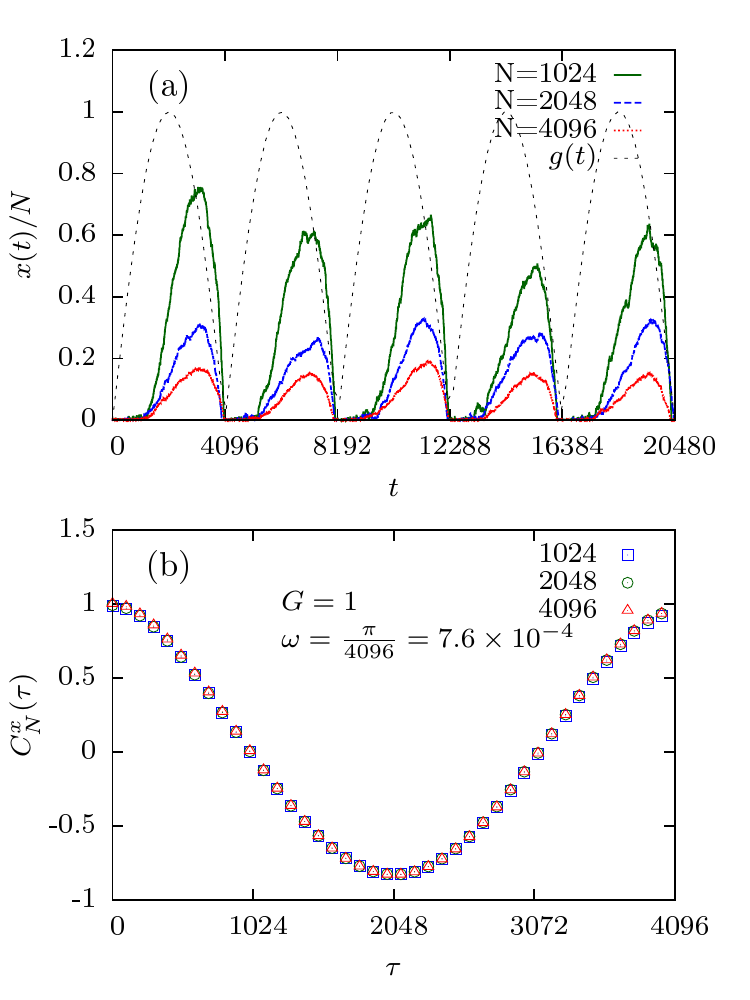}

	\caption{ \label{fig:2} (a) The end-to-end separation $x(t)$ as a
	function of time $t$ for the DNA of lengths $N=1024$, 2048 and 4096 when it is
	subjected to a periodic force of amplitude $G=1$ at frequency $\omega = \pi/4096 
	= 7.6 \times 10^{-4}$. The force variation is shown by thin dotted lines. 
	(b) { Data collapse of time autocorrelation function $C_N^x(\tau)$ of extension 
	$x$, defined in Eq. (\ref{corrdef}), for various chain lengths $N$ at same frequency and 
	force amplitude as in (a)}.}

\end{figure}
}

\newcommand{\figThree}{
\begin{figure*}[t]
	\centering
	\includegraphics[width=\textwidth]{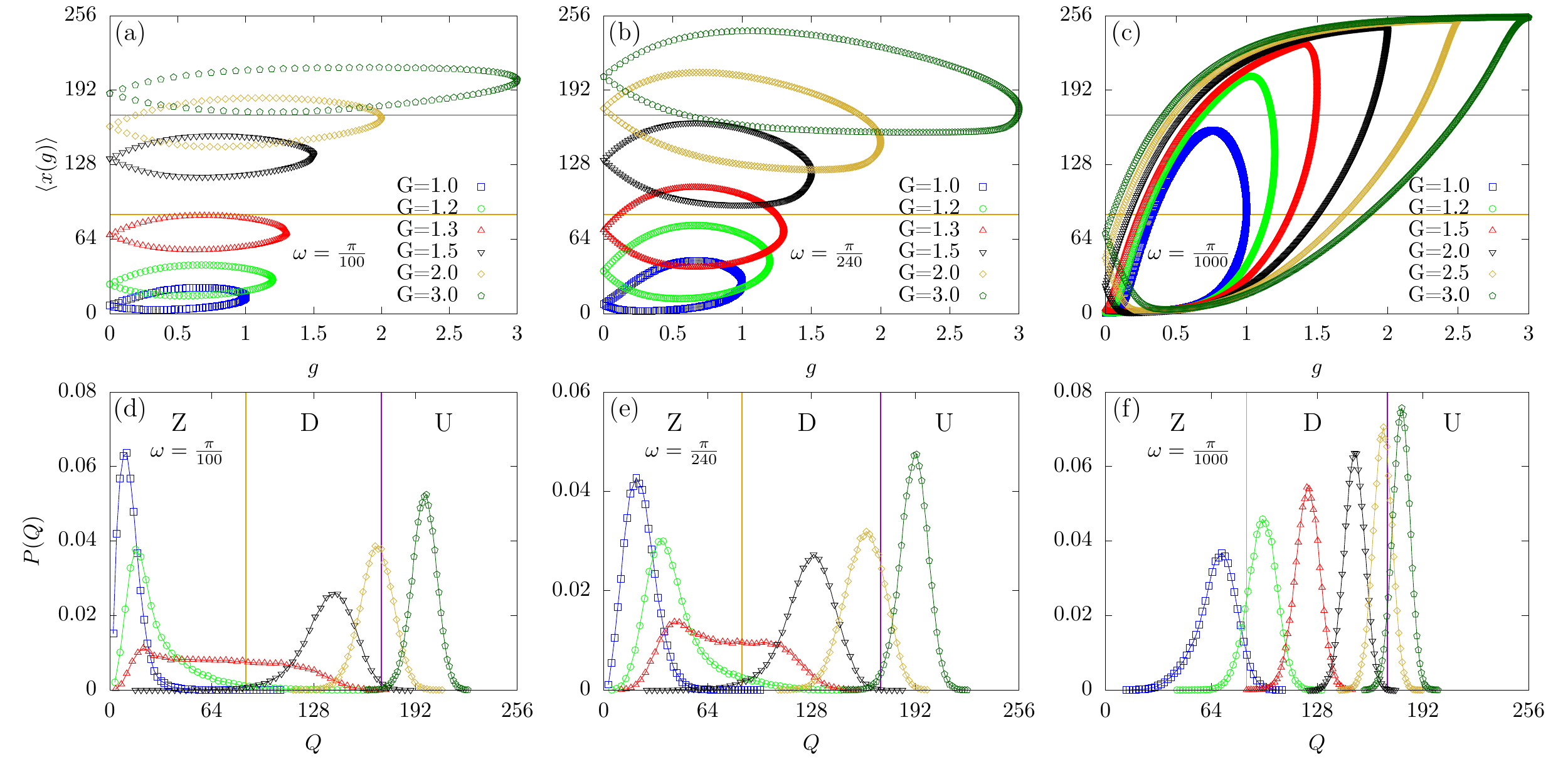}

	\caption{ \label{fig:3} (Top) Average extension $\langle x(g)
	\rangle$ as a function of force $g$ for various force amplitudes $G$ as indicated
	with frequencies of pulling (a) $\omega = \pi/100$, (b) $\omega = \pi/240$, and (c)
	$\omega = \pi/1000$. (Bottom) The normalized probability distribution $P(Q)$ 
	of order parameter $Q$ at frequencies (d) $\omega = \pi/100$, (e) $\omega = \pi/240$,
	and (f) $\omega = \pi/1000$. The length of the DNA used is $N=128$. }
\end{figure*}
}

\newcommand{\figFour}{
\begin{figure}[b]
	\centering
	\includegraphics[width=3.5in]{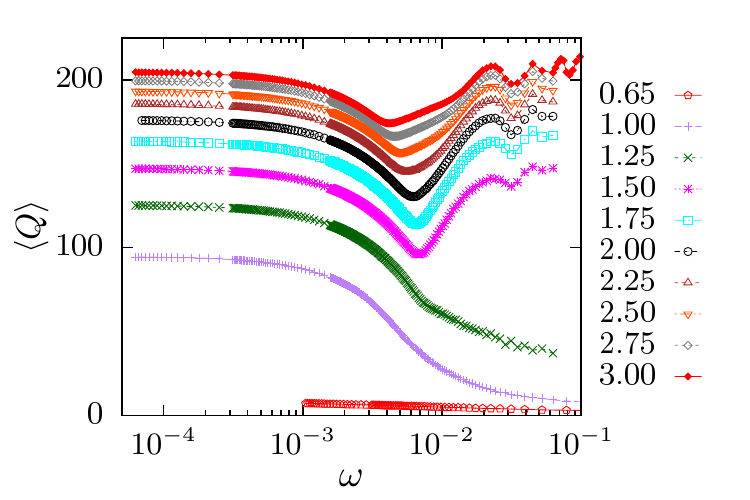}

	\caption{ \label{fig:4} Averaged dynamical order parameter $\langle Q \rangle$
	as a function of frequency $\omega$ (in log-scale) of the periodic force at various amplitudes
	$G$ for the DNA of length $N=128$. }

\end{figure}
}

\newcommand{\figFive}{
\begin{figure*}[t]
	\centering
	\includegraphics[width=7.0in]{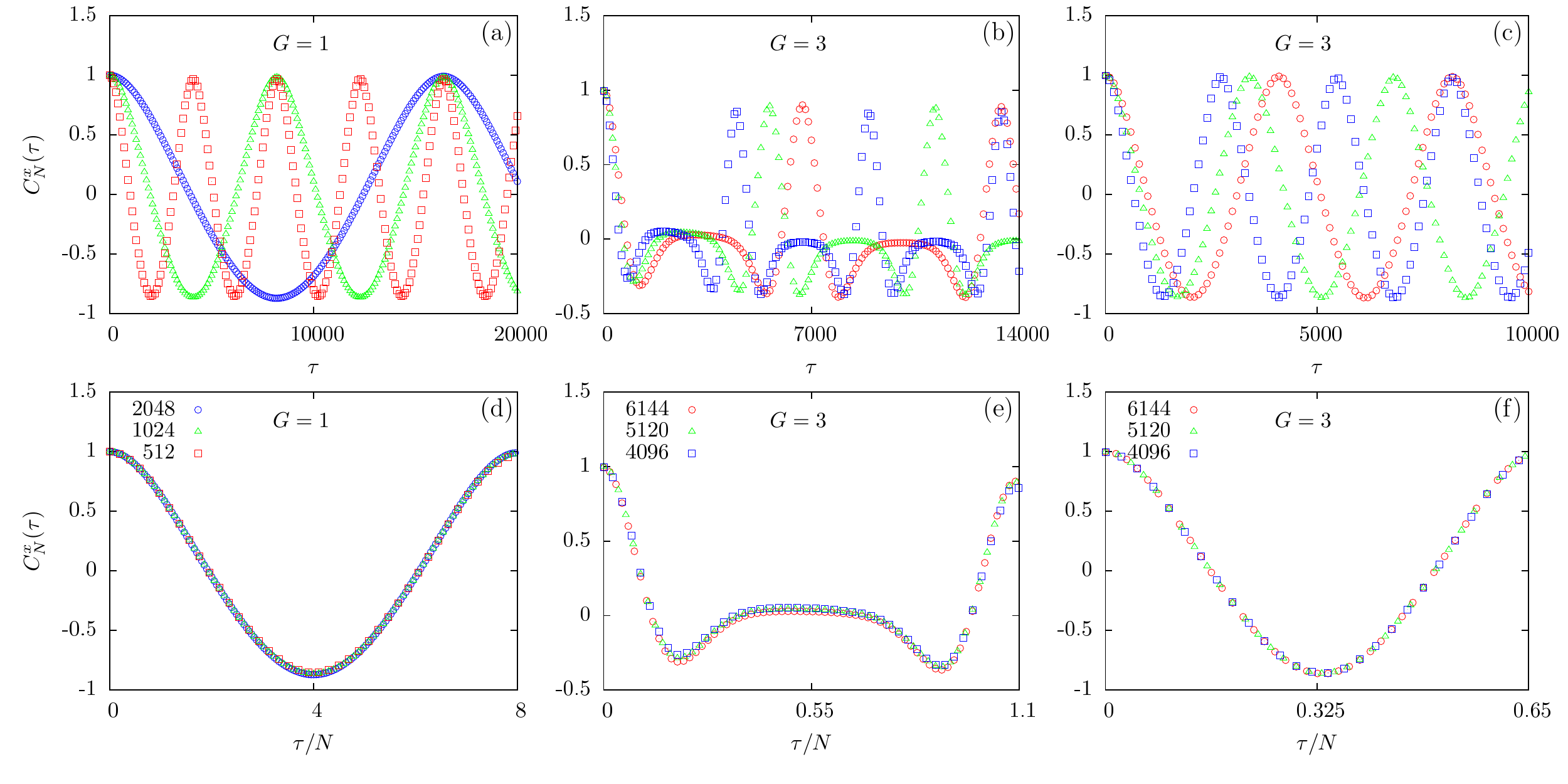}

	\caption{ \label{fig:5} (Top) The correlation function of extension 
	between the last monomers of two strands $C_N^x(\tau)$ as a function of $\tau$ for DNA of
	various lengths $N$ at force amplitudes (a) $G=1$ [(b) and (c)] $G=3$. (Bottom) The 
	data for various $N$ collapse to a single scaling curve when $\langle x(t) x(0) \rangle$ 
	is plotted with $\tau/N^{z}$ with dynamic exponent $z=1$.  }

\end{figure*}
}

\newcommand{\figSix}{
\begin{figure}[t]
	\centering
	\includegraphics[width=3.0in]{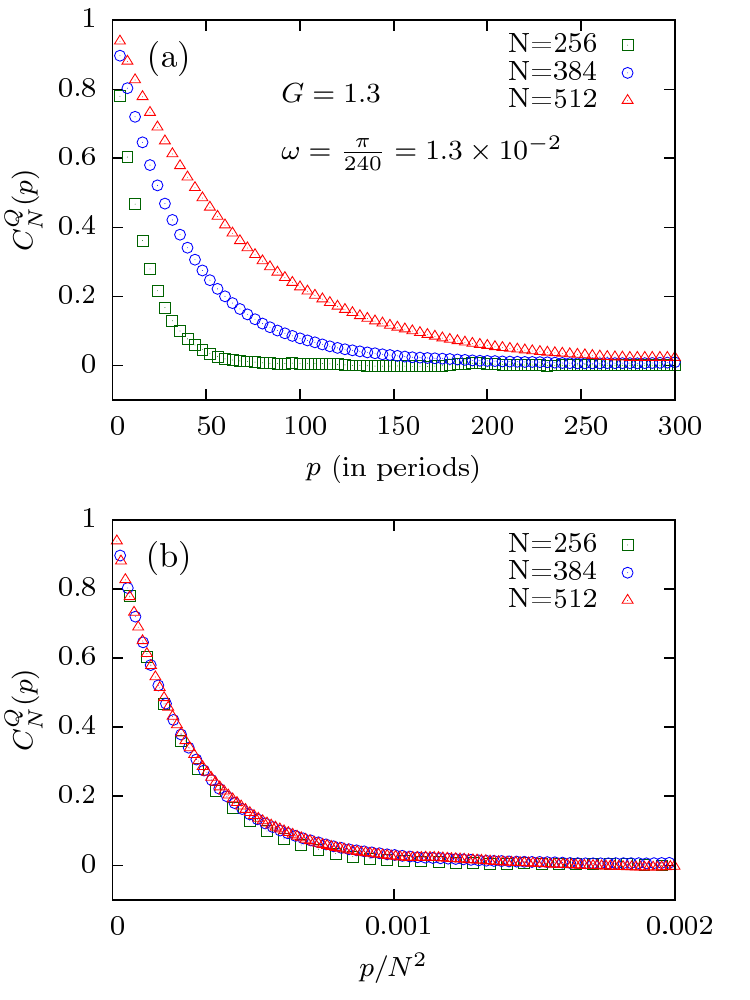}

	\caption{ \label{fig:6} (a) Normalized autocorrelation function, $C_N^Q(p)$,
    of the order parameter $Q$ for the DNA of various lengths at force amplitude $G=1.3$ and frequency
    $\omega = \pi/240$ in units of MCS. (b) Collapse of data shown in (a). }

\end{figure}
}

\newcommand{\figSeven}{
\begin{figure*}[t]
	\centering
	\includegraphics[width=5.0in]{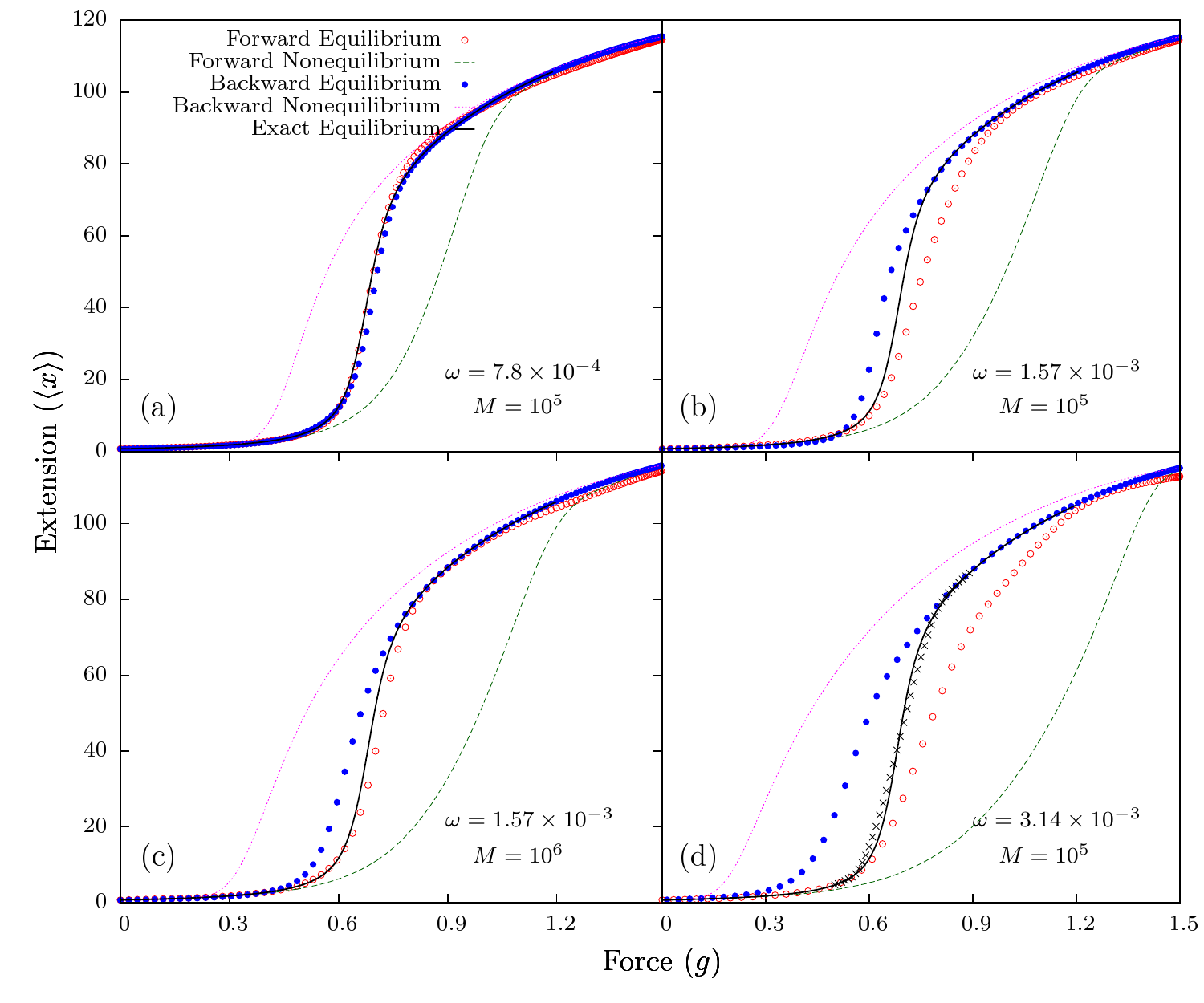}
		
	\caption{\label{fig:7} Force $g$ versus average extension $\langle x \rangle$ between
	the end monomers of the two strands of the DNA of length $N=64$ subjected to a periodic force of
	amplitude $G=1.5$ and frequencies $\omega = 7.8 \times 10^{-4}$ (a), $\omega = 1.57 \times 10^{-3}$
	[(b) and (c)], and $\omega=3.14 \times 10^{-3}$(d). In all the plots dashed (dotted) lines 
	represent the forward (backward) nonequilibrium paths and open (closed) circles represent the
	equilibrium curves obtained by taking the weighted average from $M$ [$10^6$ in (c) and $10^5$ in 
	(a), (b), and (d)] forward (backward) trajectories. The solid lines in all the plots are from the
	exact transfer matrix calculations. The symbol $\times$ in (d) at the transition region is obtained 
	by using cubic spline interpolation scheme on the data up to $g=0.55$ from the forward, and the 
	data beyond $g=0.80$ from the backward paths. }

\end{figure*}
}

\begin{document}

\title{Unzipping DNA by a periodic force: Hysteresis loops, Dynamical order parameter, Correlations 
and Equilibrium curves}

\author{M. Suman Kalyan}
\email{maroju.sk@gmail.com}

\author{Rajeev Kapri}
\email{rkapri@iisermohali.ac.in}
\affiliation{Department of Physical Sciences, Indian Institute of Science Education and
Research Mohali, Sector 81, Knowledge City, S. A. S. Nagar, Manauli PO 140306, India.}

\begin{abstract}

	The unzipping of a double stranded DNA whose ends are subjected to a time dependent periodic 
    force with frequency $\omega$ and amplitude $G$ is studied using Monte Carlo simulations. 
    We obtain the dynamical order parameter, $Q$, defined as the time average extension between 
    the end monomers of two strands of the DNA over a period, and its probability distributions 
    $P(Q)$ at various force amplitudes and frequencies. We also study the time autocorrelations 
    of extension and the dynamical order parameter for various chain lengths. The equilibrium 
    force-distance isotherms were also obtained at various frequencies by using non-equilibrium
    work measurements.

\end{abstract}


\date{\today}

\maketitle

\section {Introduction}

The unzipping of a double stranded DNA (dsDNA) is a crucial step in biological processes
like DNA replication and RNA transcription. This is achieved \textit{in vivo} by enzymes
like helicases and polymerases~\cite{Watson2003}. In the last two decades there have been
several \textit{in vitro} experimental studies
(see~\cite{Smith1996,Wang1997,Roulet1997,Bockelmann2002,Danilowicz2004,Ritort2006} and
references therein) on unzipping transitions due to the development of single molecule
manipulation techniques such as atomic force microscopy, optical and magnetic tweezers,
etc. These studies have been supplemented by theoretical
modeling~\cite{Bhattacharjee2000,Lubensky2000,Sebastian2000,Marenduzzo2001,Marenduzzo2002,Kapri2004,Kapri2006,Kapri2007,Kapri2008,Kapri2009,Kumar2010,Kalyan2015}
which have provided useful insights into the problem including the unzipping phase
diagram~\cite{Marenduzzo2002,Kapri2004}. It was found that the unzipping of a dsDNA is a
first order phase transition implying that the DNA remains in a zipped phase unless the
force that pulls apart its strands exceeds a critical
value~\cite{Bhattacharjee2000,Lubensky2000}. Above this critical force, the DNA is in the
unzipped phase in which the strands are far apart. The critical force depends on the
temperature of the surrounding and decreases to zero when the temperature becomes equal to
the melting temperature of the DNA. At this temperature the thermal denaturation of the
DNA takes place in which the strands of the DNA remains apart from each other and acquire
conformations that increases the entropy of the system.

In recent years, the unzipping studies have shifted toward the periodic forcing of DNA as
it is more closer to the phenomenon that occurs in a living cell. The unzipping of DNA
inside the cell is a nonequilibrium process initiated by motor proteins called helicases.
These motor proteins require a constant supply of energy, obtained from ATP hydrolysis,
for their functioning and apply force on the DNA in a cyclic manner~\cite{Watson2003}
(e.g., PcrA helicase~\cite{Velankar1999}). This periodic force can cause unbinding and
rebinding of biomolecules~\cite{Hatch2007,Li2007,Friddle2008,Tshiprut2009,Min2013} that can
provide useful information on the kinetics of conformational transformations, the
potential energy landscape, and can be used in controlling the folding pathways of a single
molecule~\cite{Li2007}. By applying a periodic force on the ends of the DNA, it was found
using Langevin dynamics \cite{Mishra2013,Sanjay2013,Rakesh2013,Sanjay2016,Pal2018} and
Monte Carlo~\cite{Kapri2012,Kapri2014} simulations { on the coarse grained 
models} that the dsDNA can be taken from a zipped to an unzipping phase (or {vice 
versa}) dynamically either by changing the frequency of the force and keeping the amplitude 
constant, or by changing the amplitude of the force and keeping the frequency constant. 
When the strands of the DNA are pulled away by a periodic force $g(t)$, the extension between 
the end monomers $x(t)$ follows the force with a lag. The average extension $\langle x(g) 
\rangle$ when plotted against the magnitude of force $g$ shows a hysteresis loop whose area, 
which represents the amount of energy dissipated in the system, is a dynamical order 
parameter~\cite{Chakrabarti1999}. {Hysteresis is usually associated with a
first-order phase transition due to the coexistence of two phases at a first-order phase 
boundary. These two phases are separated by an interface whose energy acts as a barrier between 
them. Near the phase boundary, there is a region of metastability where the system can stay in 
its previous phase even after crossing the phase boundary. From the dynamics point of view, 
the relaxation time or the time scale to cross the barrier becomes large near the transition, 
and therefore, there is a conflict between relaxation and the time scale of change of 
parameters, which produces hysteresis.}

For systems exhibiting dynamic phase transitions, the time average of the order parameter
(extension between the end monomers of two strands for the present case) over a time
period also serves as another dynamical order parameter (say $Q$)\cite{Chakrabarti1999}.
Although there have been many studies on periodic forcing of DNA that have focused on the
hysteresis loop area and its scaling with the force amplitude and
frequency,~\cite{Sanjay2013,Rakesh2013,Kapri2014,Sanjay2016,Pal2018} there is only one
Langevin dynamics simulation study to the best of our knowledge that focuses on the
behavior of $Q$.~\cite{Mishra2013} In this paper we discuss the variation of average
dynamical order parameter $\langle Q \rangle$ and the probability distributions $P(Q)$ as
a function of frequency and force amplitude using Monte Carlo simulations. We find that
the time autocorrelation of extension between the end monomers of two strands behaves with
length $N$ as $t/N^z$ with dynamic exponent $z=1$ and the dynamical order parameter $Q$
varies with length as $p/N^{z_p}$, where $p$ represents the number of time periods of the
force, with exponent $z_p=2$. We also obtain the equilibrium force-distance isotherms for
the DNA using the nonequilibrium work measurements on the trajectories traced by the
distance between the end monomers of two strands of the DNA due to periodic forcing.
 
The paper is organized as follows: In Sec.~\ref{sec:model}, we define our model and the
quantities of interest. In Sec.~\ref{sec:results}, we first present our results and
discuss them. We summarize our results in Sec.~\ref{sec:summary}. 

\section{Model} \label{sec:model}

The model used in this paper has been used previously to study the unzipping of DNA by periodic 
forcing.~\cite{Kapri2012,Kapri2014} In this model, the two strands of a homopolymer DNA are
represented by two directed self-avoiding walks on a ($d=1+1$)-dimensional square lattice.
The walks starting from the origin are restricted to go towards the positive direction of the 
diagonal axis ($z$-direction) without crossing each other, { i.e., in every step, the 
$z$ coordinate is incremented by 1 and the $x$ coordinate changes by $\pm 1$. The projection of 
a monomer on the $x-$axis gives its $x$ coordinate}. The directional nature of walks takes care 
of self-avoidance and the correct base pairing of DNA, i.e., the monomers that are complementary 
to each other are allowed to occupy the same lattice site. For each such overlap there is a gain 
of energy $-\epsilon$ ($\epsilon >0$).  One end of the DNA is anchored at the origin and a 
time-dependent periodic force
\begin{equation}
	\label{eq:1}
	g(t) = G \left| \sin \left( \omega t \right) \right|
\end{equation}
with angular frequency $\omega$ and amplitude $G$ acts along the transverse direction ($x$
direction) at the free end. Throughout the paper, by frequency we mean the angular frequency. 
The schematic diagram of the model is shown in Fig.~\ref{fig:1}. 

\figOne

In the static force limit (i.e., $\omega \to 0$), the model can be solved exactly via generating 
function and exact transfer matrix techniques, and has been used to obtain the phase diagrams of 
the DNA unzipping.~\cite{Marenduzzo2001,Marenduzzo2002,Kapri2004} For the static force case, the 
temperature dependent phase boundary is given by
\begin{equation}
	g_c(T) = - \frac{T}{2} \ln \lambda (z_2),
	\label{eq:gc}
\end{equation}
where $\lambda(z) = (1- 2z -\sqrt{1-4z})/(2z)$ and $z_2 = \sqrt{1 -e^{-\beta \epsilon} } - 1 +
e^{-\beta \epsilon}$. The zero force melting takes place at a temperature $T_m = \epsilon/\ln (4/3)$ 
(for details see Ref.~\cite{Kapri2012}). From Eq.~(\ref{eq:gc}), the critical force at temperature 
$T=1$, which is the temperature used in this study, is obtained as $g_c(1) = 0.6778\dots$. 
{ Although the above model ignores finer details like bending rigidity of the dsDNA,
sequence heterogeneity, stacking of base pairs, etc., it was found that the basic features, such as 
the first order nature of the unzipping transition and the existence of a re-entrant region allowing
unzipping by decreasing temperature, are preserved by this two dimensional model\cite{Marenduzzo2001,
Marenduzzo2002}}.

We perform Monte Carlo simulations of the model by using the METROPOLIS algorithm. In our model, 
the directional nature of the walks prevents the self-crossing of strands. To avoid mutual crossing 
of strands, we allow strands to undergo Rouse dynamics with local corner-flip or end-flip
moves~\cite{Doi1986} that do not violate mutual avoidance. The elementary move consists of selecting 
a random monomer from a strand, which itself is chosen at random, and flipping it. If the move 
results in overlapping of two complementary monomers, thus forming a base-pair between the strands, 
it is always accepted as a move. The opposite move, i.e., the unbinding of monomers, is chosen with 
the Boltzmann probability $\xi = \exp(- \varepsilon/ k_B T)$. { If the chosen monomer
is unbound and would remain unbound after the move is performed, it is always accepted.}  The time 
is measured in units of Monte Carlo Steps (MCSs). One MCS consists of $2N$ flip attempts, i.e., on
average, every monomer is given a chance to flip. Throughout the simulation, the detailed
balance is always satisfied. From any starting configuration, it is possible to reach any
other configuration by using the above moves.  Throughout this paper, we have chosen
$\varepsilon = 1$ and $k_B = 1$. 

At any given frequency $\omega$ and the force amplitude $G$, as the time $t$ is
incremented by unity, the external force $g(t)$ changes, according to Eq.~(\ref{eq:1}),
from $0$ to a maximum value $G$ and then decreases to $0$. Between each time increment,
the system is relaxed by a unit time (1 MCS). Upon further increment in $t$, the above
cycle gets repeated again and again.  Before taking any measurement, the simulation is run
for $2000$ cycles so that the system can reach the stationary state.  

In our simulations, we monitor the distance between the end monomers of the two strands,
$x(t)$, as a function of time for various force amplitudes $G$ and frequencies $\omega$. 
{ Note that a monomer on flipping always goes to the opposite corner of the 
square lattice in this model. Thus, the distance changes by 2 units (i.e., the length of 
the diagonal) in each flip}. From the time series $x(t)$, we can define a dynamical quantity 
$Q$ as the time average of $x(t)$ over a complete period
\begin{equation} \label{eq:Q}
	Q = \frac {\omega}{\pi} \oint x(t) dt.
\end{equation}
Following Chakrabarti and Acharyya,~\cite{Chakrabarti1999} we call $Q$ as the dynamical order 
parameter.

From the time series $x(t)$, we can also obtain the extension $x(g)$ as a function of
force $g$. Since the force is periodic in nature, one closed loop is obtained per cycle.
On averaging over various cycles, we obtain the average extension $\langle x(g) \rangle$.
If the force amplitude $G$ is not very small, and the frequency $\omega$ of the periodic
force is sufficiently high to avoid equilibration of the DNA, the average extension,
$\langle x(g) \rangle$, for the forward and the backward paths is not the same and we see
a hysteresis loop. The area of the hysteresis loop, $A_{loop}$, defined by
\begin{equation}
	A_{loop} = \oint  \langle x(g) \rangle dg,
	\label{eq:area}
\end{equation}
depends upon the frequency $\omega$ and the amplitude $G$ of the oscillating force and also
serves as another dynamical order parameter.~\cite{Chakrabarti1999}

In Ref.~\cite{Kapri2014}, we have reported the behavior of $A_{loop}$ at
high and low frequencies at various force amplitudes $G$ using Monte
Carlo simulations. In this paper, we focus mainly on the results related
to the dynamical order parameter $Q$.

\section{Results and Discussions} \label{sec:results}

\subsection{Hysteresis loops} 

\figTwo

\figThree

In Fig.~\ref{fig:2}(a), we have plotted five different cycles of the
scaled extension $x(t)/N$ as a function of time $t$ for DNA of various
lengths $N=1024$, 2048, and 4096, when it is subjected to a periodic
force of amplitude $G=1$ at frequency $\omega = \pi/4096 = 7.6 \times
10^{-4}$.  { The time required to unzip the DNA is directly proportional
to its length. Since the magnitude of the force continues to increase
much beyond the critical force $g_c$, the unzipped section of the DNA
keeps on stretching. For a given frequency and force amplitude, only a
finite length of the chain could be unzipped, which is directly
proportional to the extension $x(t)$ due to the geometry of the problem.
To plot extensions for various chain lengths on the same scale, we
divide it by the chain length $N$. It is easy to see that the scaled
extension between the strands depends on the chain length.} The shorter
chain lengths have larger scaled extension and {vice versa}.  To get the
same scaled extension between the strands of a longer chain at a fixed
force amplitude, $G$, one needs to decrease the frequency $\omega$ of
the pulling force { as it will provide more time to the chain to relax
at force values above $g_c$. By using a simple analysis, it was found in
Ref.~\cite{Kapri2014} that at low frequencies the hysteresis loop area
behaves as $G^{\alpha}\omega^{\beta}$, with $\alpha=1$ and $\beta=5/4$,
reaches a maximum at $\omega^{\star}$ and then decreases as $1/\omega$
at high frequencies. The frequency $\omega^{\star}$ depends on the force
amplitude $G$ and is inversely proportional to the chain length $N$
showing that the strands of DNA could only be opened by taking the
frequency $\omega \to 0$ (i.e., the static limit) in the thermodynamic
limit (i.e., $N\to \infty$).} Note that the amplitude $G$ of the
periodic force should always be greater than the critical force needed
to unzip the dsDNA at that temperature, given by Eq.~(\ref{eq:gc}).
Otherwise, the strands of the DNA will remain zipped, with scaled
separation $x(t)/N \to 0$, even by decreasing the frequency of the
force.  It is also possible to unzip the dsDNA by keeping the frequency
fixed and increasing the amplitude $G$ of the force. 

The average extension between the strands $\langle x(g) \rangle$ as a
function of force $g$ when DNA of length $N=128$ is pulled by a periodic
force of various amplitudes, $G$, ranging from $G=1$ to $G=3$, for three
different frequencies $\omega = \pi/100$, $\pi/240$, and $\pi/1000$ is
shown in Fig.~\ref{fig:3}(a)-\ref{fig:3}(c). Following Mishra \textit{et
al.,}\cite{Mishra2013} we divide the extension in three different
regions for identifying the phases of the DNA shown by thin solid lines.
For a DNA of length $N=128$, the maximum allowed extension between the
strands of the DNA, due to the structure of the lattice, is $x_{\rm max}
= 2N = 256$. Therefore, when the extension $x$ is less than one-third of
$x_{\rm max}$ (i.e., $x \le x_{\rm max}/3$), we assign the DNA to be in
the zipped ($Z$) state. If the extension is more than two-thirds of
$x_{\rm max}$ (i.e., $x \ge 2x_{\rm max}/3$), the DNA is assigned to be
in the unzipped ($U$) state, and when $x_{\rm max}/3 < x < 2x_{\rm
max}/3$, the DNA is assumed to be in between the zipped and the unzipped
state, henceforth called a dynamic ($D$) phase. 

At a higher frequency $\omega = \pi/100$ [Fig.~\ref{fig:3}(a)], the
external force changes very rapidly and the DNA gets no time to relax;
hence we get a hysteresis loop of small area.  For lower values of force
amplitudes ($G=1.0$ to 1.3), the average extension between the strands
at force value $g=0$ (i.e., $\langle x(0)\rangle$), is very small, which
indicates that at these force amplitudes, the DNA is in zipped
configuration and the stationary state is a zipped state ($Z$). As the
amplitude increases, so does the value of $\langle x(0)\rangle$ but the
area of the loop still remains small. For $G=3$, the value of $\langle
x(0)\rangle$ is more than $2 x_{\rm max}/3$, which indicates that the
DNA is in the unzipped configuration and the stationary state of the DNA
is an unzipped state ($U$). On decreasing the frequency slightly, i.e.,
$\omega = \pi/240$ [Fig.~\ref{fig:3}(b)], the DNA now gets slightly more
time to relax and the area of loop increases. The stationary states for
the smaller and higher $G$ values remain same as that of
Fig.~\ref{fig:3}(a). On decreasing the frequency further, i.e., $\omega
= \pi/1000$ [Fig.~\ref{fig:3}(c)], we see that the DNA now gets more
time to relax under the influence of an external force which is seen by
the increased loop area. Even at this frequency, the DNA could not be
completely unzipped for smaller force amplitudes ($G=1.0$, 1.2 and 1.5),
hence the loop area is still smaller. However, for higher amplitudes
($G= 2.0$, 2.5 and 3.0), the DNA gets completely unzipped with a larger
hysteresis loop area.  

\figFour

\subsection{Dynamical order parameter}

The dynamical order parameter averaged over $10^6$ cycles, $\langle Q
\rangle $, is plotted as a function of frequency $\omega$ (in log-scale)
for the DNA of length $N=128$ at various force amplitudes $G$ in
Fig.~\ref{fig:4}. For amplitude $G=0.65$, the maximum value of the
periodic force is always less than the critical force (i.e., $g_c(1) =
0.6778\dots$) needed to unzip the DNA at $T=1$, hence the DNA remains in
the zipped state irrespective of the frequency of the periodic force. As
discussed in Subsection\ref{sec:results}A, the stationary state of the
DNA is a zipped ($Z$) state when the force amplitude $G$ is 1.0 and
1.25, as could be seen by lower value of $\langle Q \rangle$ at higher
frequencies. As at such frequencies, the DNA does not get time to
respond to the change in the force value and effectively remains in its
stationary state, which is a zipped configuration in this case. However,
as the frequency decreases, the DNA starts responding to the periodic
force and the value of $\langle Q \rangle$ increases and gets saturated.
When the force amplitude is very high (i.e., $G=3$), the stationary
state of the DNA is an unzipped ($U$) state as seen in the plot by the
maximally allowed $\langle Q \rangle$ value at higher frequencies. As
the frequency decreases, the value of $\langle Q \rangle$ shows
oscillations before becoming constant at lower frequencies. The
frequency at which there appears a larger dip in $\langle Q \rangle$ is
exactly the same at which the hysteresis loop area, $A_{\rm loop}$,
shows a maximum. The frequencies at which there are other minima in
$\langle Q \rangle$ are also similar to the frequencies at which
secondary maxima occur in $A_{\rm loop}$.~\cite{Kapri2014}

From the above discussion, we see that $\langle Q \rangle $ does not
give any further useful information that has not already been obtained
from the hysteresis loop area.~\cite{Kapri2014} Therefore, we study the
probability distributions, $P(Q)$, of the dynamical order parameter $Q$
defined by Eq.~(\ref{eq:Q}). We find that the allowed values of $Q$ do
not follow a regular pattern and appears randomly. The distributions
$P(Q)$ are obtained by binning $Q$ values acquired in $10^6$ cycles of
the periodic force. The normalized distributions are shown in
Figs.~\ref{fig:3}(d)-\ref{fig:3}(f) for the DNA of length $N=128$ at
various frequencies $\pi/100$, $\pi/240$ and $\pi/1000$, for which the
hysteresis loops are shown in Figs.~\ref{fig:3}(a)-\ref{fig:3}(c). At a
higher frequency $\omega = \pi/100$ [Fig.~\ref{fig:3}(d)], the
distributions $P(Q)$ for lower values of amplitudes $G=1.0$ and 1.2 are
sharply peaked at lower values of $Q$ showing that the DNA is in the
zipped ($Z$) phase.  At an amplitude of $G=1.3$, the distribution $P(Q)$
becomes broader and spans both the zipped ($Z$) and dynamic ($D$)
phases. On increasing the amplitude further  (i.e., $G=1.5$), the
distribution becomes narrower again with a peak for intermediate $Q$
values that lies in the dynamic $D$ phase. On increasing the amplitude
further ($G=3$), the distribution is again sharply peaked at higher $Q$
values showing that the DNA is in the unzipped ($U$) phase. For a
slightly lower frequency $\omega = \pi/240$ [Fig.~\ref{fig:3}(e)], the
distributions are qualitatively similar to Fig.~\ref{fig:3}(d) but with
a slightly more pronounced double peak structure for $G=1.3$. However,
at much lower frequencies $\omega = \pi/1000$ [Fig.~\ref{fig:3}(f)], the
distributions at all $G$ values become sharp. From these distributions,
we can clearly see that the dsDNA could be taken from a zipped ($Z$) to
an unzipped ($U$) state via a dynamic ($D$) state or {vice versa} at a
constant frequency by changing the force amplitude. For higher
frequencies, the distributions are very sensitive to the value of force
amplitude $G$. There are regions where a small change in the value of
$G$ could change a sharp distribution to a broader one. However, we
could not get a three peak structure as seen in Langevin dynamics
simulation study of shorter DNA hairpin under periodic
force.~\cite{Mishra2013}

\subsection{Correlation functions}

In this section, we study the behavior of correlation functions of the
extension between the end monomers of the two strands of the DNA and the
dynamical order parameter $Q$ as a function of time.  

\figFive

\subsubsection{Extension}

In the presence of a periodic force, the DNA undergoes a transition from
a zipped to an unzipped phase in each cycle. On a square lattice, with
unit lattice spacing, the end-to-end displacement for the DNA having $N$
monomers can vary between $N\sqrt{2}$, for a completely stretched
configuration, and $N$ for a zigzag configuration that has maximum
entropy. The later configuration is taken by the DNA for lower force
values where it is in the zipped state.  However, for higher force
values, the DNA is in a completely stretched unzipped state having
maximum length. Therefore, in the presence of a periodic force, the
length of the DNA fluctuates and have longitudinal modes. Furthermore,
due to the geometry of the square lattice, the change in length of the
DNA by flipping a monomer (i.e., along the $z$ axis) is exactly equal to
the change in the separation of the end monomers (i.e., along the $x$
axis). Therefore, the length correlation function is exactly equal to
the correlation function for the extension between the end monomers of
the two strands. 

We define the normalized time autocorrelation function of extension $x$ between the strands 
of DNA of length $N$ as
\begin{equation}\label{corrdef}
C^{x}_{N}(\tau) = \frac{\langle x(t) x(t + \tau) \rangle - \langle x(t) \rangle^2 } 
	{ \langle x(t)^2 \rangle - \langle x(t) \rangle^2 }.
\end{equation}
The correlation function of extension $C^x_N(\tau)$ as a function of time $\tau$ for the DNA of 
various lengths $N=1024$, 2048, and 4096 when it is subjected to a periodic force of frequency 
$\pi/4096$ at force amplitude $G=1$ is plotted in Fig.~\ref{fig:2}(b). A nice collapse for various
lengths suggests that they have similar correlation, independent of DNA length, when subjected to 
a periodic force of same frequency. The extension between two times that differ by $\tau$ is 
maximally correlated (or anticorrelated), i.e., $C^x_N(\tau)= 1 \ (-1)$, when $\tau$ is an 
integral (half-integral) multiple of the time period of the oscillating force. In between,
the correlation first decreases from a maximum to a minimum as $\tau$ increases from integral to 
half-integral multiple of the time period and then increases again to reach the maximum when $\tau$ 
increases from half-integral to integral multiple of the time period. 

It is interesting to calculate the correlations of extension,
$C_N^x(\tau)$, at different frequencies.  In
Figs.~\ref{fig:5}(a)-\ref{fig:5}(c), we have plotted $C^x_N(\tau)$ as
function of time $\tau$ for the DNA of various lengths, $N$, at force
amplitudes $G=1$ and $3$. The time required to unzip the DNA is
proportional to its length. Therefore, on increasing the length of the
DNA, we need to reduce the frequency of the applied force to keep the
product $\omega N$ constant. By using a simple analysis for the
condition of maximum hysteresis loop, it was found~\cite{Kapri2014} that
for $G=1$, the frequency $\omega$ of the applied force and the length
$N$ of the DNA satisfy the expression $\omega = \pi/8N$.  This relation
is used to fix the frequency of various chain lengths in
Fig.~\ref{fig:5}(a). For higher force amplitudes (e.g., $G=3$), it was
found that the area of the hysteresis loop and the average dynamical
order parameter show an oscillatory behavior. At the location of first
minimum and the second maximum of the hysteresis loop area, the relation
between $\omega$ and $N$ becomes $\omega = 11 \pi/ 12 N$ and $\omega = 3
\pi/2N$, respectively. We use the above relations to fix the frequency
of various chain lengths in our simulations. We have plotted
$C^x_N(\tau)$ as a function of $\tau$ for $N=1024, 2048$, and $4096$ in
Figs.~\ref{fig:5}(b) and \ref{fig:5}(c). When the above data for various
$N$ are plotted as a function of $\tau /N$, we obtain an excellent
collapse giving the dynamical exponent $z=1$ ($\tau \sim N^{z}$). The
data collapse are plotted in Figs.~\ref{fig:5}(d)-\ref{fig:5}(f).

\subsubsection{Dynamical order parameter}

\figSix

We define the normalized time-autocorrelation function of the dynamical order parameter $Q$ for 
length $N$ as
\begin{equation}
C_N^Q(p) = \frac{ \langle Q(i) Q(i+p) \rangle - \langle Q(i) \rangle^2 }
	{ \langle Q(i)^2 \rangle - \langle Q(i) \rangle^2 },
\end{equation}
where $p$ represents the number of time periods. In Fig.~\ref{fig:6}(a), we have plotted $C_N^Q(p)$ 
as a function of $p$ for the DNA of lengths $N=256$, 384, and 512 when it is subjected to a periodic 
force of amplitude $G=1.3$ at frequency $\omega = \pi/240$. The increase in the correlation time with
increasing system sizes gives evidence of critical slowing down of the system, providing support 
for the existence of a dynamical phase transition. When $C_N^Q(p)$ for various lengths are plotted as 
a function of $p/N^2$ (Fig.~\ref{fig:6}(b)), we get a nice collapse, implying that the number of time
periods $p$ after which the order parameter $Q$ becomes completely uncorrelated depends on the system 
size as $p \sim N^{z_p}$, with dynamic exponent $z_p=2$.

\subsection{Equilibrium curves}

\figSeven

Kapri~\cite{Kapri2012} has recently developed a procedure in the fixed force ensemble to
obtain the equilibrium force-distance isotherms using nonequilibrium measurements. The
scheme is similar to that of Hummer and Szabo~\cite{HummerPNAS2001} in the fixed velocity
ensemble which has been used successfully to obtain the zero force free energy on single
molecule pulling experiments~\cite{GuptaNPhys2011}. { Both these procedures use
the Jarzynski equality~\cite{Jarzynski1997}--- a work-energy theorem, which connects the 
thermodynamic free energy 
differences, $\Delta F$, between the two equilibrium states and the irreversible work done 
$W$ in taking the system from one equilibrium state to a nonequilibrium state with similar 
external conditions as that of other equilibrium state. The relation satisfied by $\Delta F$ 
and $W$ is $e^{\Delta F/ k_B T} = \langle e^{-W / k_B T} \rangle$, where $k_B$ and $T$ are 
the Boltzmann constant and the absolute temperature, respectively. The angular bracket 
denotes average over \textit{all possible paths} between the two states, which is dominated 
by rare paths.} In the study of Kapri,~\cite{Kapri2012} the two strands of the DNA 
are pulled apart by a force that is incremented by a constant rate from an initial value to 
some final value that lies above the phase boundary and then decreased back to the same 
initial value. For the present problem, the rate of change of force is not constant but we 
find that the same procedure could be used to obtain the equilibrium curves. 

In our simulation, the time is incremented in discrete steps that is measured in units of MCS.
Let $\tau_p$ denote the time period of the external force, we have $\omega = \pi / \tau_p $.
Therefore, in each cycle, the force given by Eq.(\ref{eq:1}) also changes in discrete steps as
\begin{equation}
	g_k = G | \sin (k \pi / \tau_p)|, 
\end{equation}
where $k$ can take integer values $0, 1, \dots \tau_p$. In the first half of the cycle ($k=0,
\dots \tau_p /2$), the force $g_k$ changes from the minimum value $g_k = 0$ to the maximum
value $g_k = G$. We identify this as the forward path, and in the second half of the cycle
$k=\tau_p/2 + 1, \dots \tau_p$  the force changes from $G$ to $0$ and is identify as the
backward path. 

Let $i$ and $k$ represent the indices for the cycle and the force,  respectively. The
irreversible work done over the $i$th cycle is given by
\begin{equation}
	W_{ik} = - \sum_{j=0}^{k} (g_{j+1} - g_{j}) x_{ij}, 
\end{equation}
where, $x_{ij}$ is the extension between the end monomers of the dsDNA during $i$th cycle at
force value $g_{j}$. By using $\exp(-\beta W_{ik})$ as the weight for path $i$, the equilibrium 
separation between the end monomers of the dsDNA, $x_{k}^{\rm eq}$, at force $g_k$ can
then be obtained by~\cite{SadhukhanJPA2010}
\begin{equation}
    \label{Eq:eqlm}
    x_{k}^{\rm eq} =  \frac{  \sum_{i=1}^{M} x_{ik} \exp \left( - \beta W_{ik}
    \right) } { \sum_{i=1}^{M} \exp \left( - \beta W_{ik} \right) }.
\end{equation}

We use multiple histogram technique~\cite{FerrenbergPRL1989} to approximate the density of
states which then can be used to estimate the equilibrium separation $x_{k}^{\rm eq}$. To 
achieve this, we first build up the histogram $H_{k}(x)$ at each force value $g_k$. If the 
extension at $i$th cycle
is $x$, the corresponding histogram value is incremented by $\exp(-\beta W_{ik})$
\begin{equation}
	H_k(x) = \sum_{i=1}^{M} e^{-\beta W_{ik}} \delta_{x, x_i},\
\end{equation}
where $\delta_{x,x_i} = 1$ if $x = x_i$, and zero otherwise. At each force value $g_k$, the
partition function $\mathcal{Z}_k$, obtained by using multiple histogram
technique,~\cite{FerrenbergPRL1989} is
\begin{equation}
	\mathcal{Z}_k = \sum_{x} \rho(x) e^{\beta g_k x},
\end{equation}
where the  density of states, $\rho(x)$, is given by
\begin{equation}
	\rho(x) = \slfrac{ \sum_j \frac{H_j(x)}{\mathcal{Z}_j} } { \sum_j \frac{ e^{\beta g_j
	x}}{\mathcal{Z}_j}}.
\end{equation}
These equations need to be evaluated self consistently till they converge. The equilibrium
separation $x_k^{\rm eq}$ at force value $g_k$ is then calculated by using the density of
states $\rho(x)$ as
\begin{equation}
 	x_{k}^{\rm eq} = \frac{1}{\mathcal{Z}_k} \sum_x x \rho(x) e^{\beta g_k x}.
\end{equation}

In Fig.~\ref{fig:7}, we have shown the force $g$, versus extension $\langle x \rangle$ curves
for the DNA of length $N=64$ at force amplitude $G=1.5$ and three different frequencies $\omega
= 7.8 \times 10^{-4}$, $1.57 \times 10^{-3}$, and $3.14 \times 10^{-3}$. At these frequencies,
the dsDNA gets enough time to relax so it is in a completely unzipped phase at the maximum
force value $g=1.5$ and in a completely zipped phase when the force value becomes zero.
Therefore, we can use both the forward and the backward paths to calculate the equilibrium
extension between the end monomers of the two strands. In these plots, the averaged
nonequilibrium forward (backward) paths are shown by dashed (dotted) lines. { 
The equilibrium curve is shown by a solid line. Thanks to the directed nature of the model, 
we could write a recursion relation for the partition function which could be solved 
numerically using the exact transfer matrix technique to obtain the partition function of DNA 
of length $N$ and the equilibrium force-distance isotherm.~\cite{Kapri2012} This allows us to
compare the results from our procedure to the exact result.} The forward
(backward) equilibrium extensions obtained by using the above procedure are shown by unfilled
(filled) circles. Figure \ref{fig:7}(a) shows the results for the frequency $\omega = 7.8 \times
10^{-4}$. At this frequency, the external force acting on the end monomers of the DNA changes
slowly. The DNA gets more time to relax and therefore the area of the hysteresis curve traced
by the averaged forward and backward paths of cycles is small. The equilibrium curve obtained
by using the above procedure on $M=10^5$ forward and backward paths matches excellently with the
curve obtained by using the exact transfer matrix. In Fig.~\ref{fig:7}(b), the results are
shown for frequency $\omega = 1.57 \times 10^{-3}$ that is twice the frequency used in
Fig.~\ref{fig:7}(a). The DNA gets lesser time to relax and therefore the area of the hysteresis
curve increases. On using the above procedure for $M=10^5$ cycles we get the equilibrium curves
that match with the exact equilibrium curve at lower and higher force values but not in the
intermediate region where the transition takes place. This is due to the poor statistics in the
transition region. However, if the number of cycles over which the histograms are taken are
increased we get better statistics and { the equilibrium curves obtained by using
the forward and the backward paths will tend to coincide with the exact equilibrium curve, 
which can be seen in Fig.~\ref{fig:7}(c) where $M=10^6$ cycles are used keeping the frequency 
same as in Fig.\ref{fig:7}(b).} If the frequency is doubled further (i.e., $\omega = 3.14 
\times 10^{-3}$), the DNA gets much lesser time to relax and the area of the hysteresis loop
increases. The equilibrium curves obtained by using the above procedure with $M=10^5$ cycles 
{ deviates from the actual equilibrium curve due to very poor statistics near the
transition region. To get better statistics either one has to use more cycles, as mentioned 
earlier, or use special algorithms to generate rare samples, as these rare paths have the
dominating contributions.} However, on analyzing these curves closely, we find that the
equilibrium curve obtained by using the forward (backward) path has good match with the
exact equilibrium curve in the region that lies below (above) the critical force. This led to
ask a question: can we combine the forward and the backward paths to obtain the equilibrium
curve that matches with the exact curve even in the transition region? The answer to this 
question is in the affirmative. By using an interpolation scheme on the forward and the 
backward paths, we can obtain the equilibrium curve. To demonstrate this, we use data up 
to $g=0.55$ from the forward, and the data beyond $g=0.80$ from the backward paths with 
cubic spline interpolation scheme to obtain the equilibrium curve in the transition region. 
The result shown in Fig.~\ref{fig:7}(d) by the symbol $\times$ matches reasonably well 
with the exact curve obtained using transfer matrix in the transition region.

\section{Conclusions} \label{sec:summary}

In this paper, we have reported the results of a periodically driven DNA using Monte Carlo
simulations. We have obtained the average extension between the end monomers of the strands 
as a function of force, $\langle x(g) \rangle$, for various frequencies. If the frequencies 
are not small enough, the system does not get enough time to relax and $\langle x(g) \rangle$
shows hysteresis whose area gives the amount of energy dissipated to the system. It is 
observed that the steady state configuration of the DNA at higher frequencies and lower force
amplitudes is a zipped ($Z$) state. At higher frequencies and higher force amplitudes, the 
steady state configuration of the DNA is two single strands that are far apart, i.e., the
unzipped ($U$) state. We also obtained the average dynamic order parameter $\langle Q \rangle$ 
as a function of frequency and found that it does not reveal any new information that is not
already known. Therefore, we obtained the probability distributions of $Q$ at different 
frequencies and force amplitudes. We observed that at a higher frequencies and for a small 
range of force amplitudes, the distribution $P(Q)$ is broad and spans both the zipped ($Z$) and 
the dynamic ($D$) phases. For lower and higher values of amplitudes, the probability
distributions are sharply peaked that lie in one of the phases. We have obtained the 
autocorrelations of the extension between the end monomers of two strands, $C_N^x(t)$, as a
function of time at different force amplitudes for various chain lengths. We found that the
correlation $C_N^x(t)$ scales as $t/N$ at all amplitudes. We have also obtained the
autocorrelation function of the dynamic order parameter, $C_N^Q(p)$, as a function of period
number $p$ for force amplitude and frequency at which we have a broader distribution 
$P(Q)$ for various chain lengths. We observed that the quantity $Q$ appears randomly and its 
autocorrelation function behaves as $C_N^Q(p) \sim p/N^2$. Finally, we obtained the equilibrium 
force-extension curves at different frequencies by using the nonequilibrium work measurements.
We find that it is possible to obtain the transition region with a good accuracy even for 
higher frequencies by interpolating the data of the forward and backward paths from the 
region where they are more accurate. { In our opinion, the magnetic tweezers, which
works in the fixed force ensemble, are the most suitable single molecule manipulation 
experimental technique to study the dynamical transitions in DNA unzipping with periodic 
forcing. The range of relevant frequencies depend on the length of the dsDNA and the force 
amplitude, which could be estimated by obtaining the time required in unzipping or rezipping 
of DNA at these force values. The procedure discussed in this paper would find application 
in obtaining the equilibrium curves in such experiments.}

\section*{Acknowledgements}

We thank A. Chaudhuri and D. Dhar for discussions and S. M. Bhattacharjee for his valuable 
comments on this manuscript. We acknowledge the HPC facility at IISER Mohali for generous 
computational time.

\end{document}